\documentclass[dvips,twoside,fleqn]{article}
\usepackage{times}
\usepackage{amstext}
\usepackage{amssymb}
\usepackage{espcrc2}
\usepackage{graphicx}
\title{SU(2) vacuum dynamics in applied external magnetic field}
\author{
Paolo Cea and 
Leonardo Cosmai\address{INFN - Sezione di Bari - Via Amendola 173 - 70126 Bari - Italy}}

\begin{document}

\begin{abstract}
The vacuum dynamics of SU(2) lattice gauge theory is studied by means of a gauge-invariant
effective action, both at zero and finite temperature. Working with lattices up to $32^4$ 
we check the scaling of the energy density with the magnetic length. We find that the 
screening at zero temperature of the applied external magnetic field weakens by increasing 
the temperature. 
\end{abstract}
\maketitle

\section{INTRODUCTION}

The aim of this work is to study the non-perturbative vacuum dynamics 
of the SU(2) l.g.t. in presence of a static external applied field.
To this purpose we use a gauge-invariant effective action
for the external background field  defined by means of the 
lattice Schr\"odinger functional\cite{Rossi1980,Luscher1992}
\begin{equation}
\label{Zlatt}
{\mathcal{Z}} \left[ \vec{A}^{\text{ext}} \right] =
\int {\mathcal{D}}U \exp(-S_W) \,,
\end{equation}
where $S_{\text{W}}$ is the Wilson action and 
$\vec{A}^{\text{ext}}(\vec{x}) =  \vec{A}_a^{\text{ext}}(\vec{x})\lambda_a/2$ 
is the external field.
We adopt periodic boundary conditions in space and time direction with
the integration constraint over the lattice links
\begin{eqnarray}
\label{constraints}
\lefteqn{U_\mu(x)|_{x_4=0} = U_\mu^{\text{ext}}(\vec{x},0)}
\nonumber \\
\lefteqn{= {\text{P}} \exp \left\{ + iag  \int_0^1 dt \,
A_\mu^{\text{ext}}(x+ at {\hat{\mu}}) \right\}
\,.}
\end{eqnarray}
The Schr\"odinger functional is invariant under arbitrary lattice
gauge transformations of the boundary links.
The lattice effective action for the background field
$A_\mu^{\text{ext}}(\vec{x})$ is defined by means of
the  lattice Schr\"odinger functional Eq.(\ref{Zlatt})  
(T extension in Euclidean time):
\begin{equation}
\label{effact}
\Gamma\left[ \vec{A}^{\text{ext}} \right] = -\frac{1}{T}
\ln \left\{ \frac{{\mathcal{Z}}[U^{\text{ext}}]}{{\mathcal{Z}}[0]} \right\}  \,.
\end{equation}
${\mathcal{Z}}[0]$ is the  
lattice Schr\"odinger functional
without external background field  
($U_\mu^{\text{ext}} =1$).
In the continuum  limit ($T\rightarrow\infty$)
$\Gamma[ \vec{A}^{\text{ext}}]$ is the 
vacuum energy in presence of the background
field $\vec{A}^{\text{ext}}(\vec{x})$.
Our gauge-invariant effective action can be used for a
non-perturbative
investigation of the properties of the SU(2) l.g.t. vacuum.

We consider static background fields that give rise to 
constant field strength. 
In this case 
$\Gamma[ \vec{A}^{\text{ext}} ]$
is proportional to the spatial volume 
$V$ and the relevant quantity is the 
density of the effective action:
\begin{equation}
\label{energydensity}
\varepsilon\left[ \vec{A}^{\text{ext}} \right] =
-\frac{1}{\Omega} \ln \left[
\frac{{\mathcal{Z}}[U^{\text{ext}}]}{{\mathcal{Z}}(0)} \right] 
\,, \quad \Omega=V \cdot T \,.
\end{equation}
In particular we consider an 
external constant abelian magnetic field:
\begin{equation}
\label{su2cont}
\vec{A}_a^{\text{ext}} = \delta_{a,3} \vec{A}^{\text{ext}} \,,
A^{\text{ext}}_k = \delta_{k,2} x_1 H \,,
F^{a}_{12} = \delta^{a,3} H \,.
\end{equation}
On the lattice
\begin{eqnarray}
\label{su2lat}
\lefteqn{U^{\text{ext}}_2(x) = \cos(\frac{agHx_1}{2})  + i \sigma^3 
 \sin(\frac{agHx_1}{2})}
\nonumber \\
\lefteqn{U^{\text{ext}}_1(x) =  U^{\text{ext}}_3(x) = U^{\text{ext}}_4(x) = 1 
\,.}
\end{eqnarray}
Periodic boundary conditions imply the
quantization of the magnetic field:
\begin{equation}
\label{quant}
\frac{a^2gH}{2} =\frac{2 \pi}{L_1} n_{\text{ext}} \, , \qquad  
n_{\text{ext}} \quad  {\text{integer}}   \,.
\end{equation}
We want to compute the vacuum energy density Eq.(\ref{energydensity}). 
To avoid the problem of dealing with a partition function we focus on 
$\varepsilon^{\prime}[\vec{A}^{\text{ext}}]$, the
derivative of $\varepsilon[\vec{A}^{\text{ext}}]$  with respect to
$\beta$, by taking $n_{\text{ext}}$ (i.e. $gH$) 
fixed\cite{Cea}.
$\varepsilon[ \vec{A}^{\text{ext}}]$ can be obtained by a numerical
integration in $\beta$.

We performed lattice simulations on $32 \times L^2 \times 32$ lattices 
(L=6,8,10,16,20,24,32) with periodic boundary conditions using the 
Quadrics/QH1 in Bari.  
The links belonging to the time slice $x_4=0$ and to the spatial boundary
are constrained according to Eq.(\ref{su2lat}).
The constraint on the links starting from sites belonging to the spatial boundary  
corresponds in the continuum to the usual requirement that the  fluctuations over the 
background field vanish at the spatial infinity. 

\begin{figure}[t]
\label{fig1}
\begin{center}
\includegraphics[width=7.5cm]{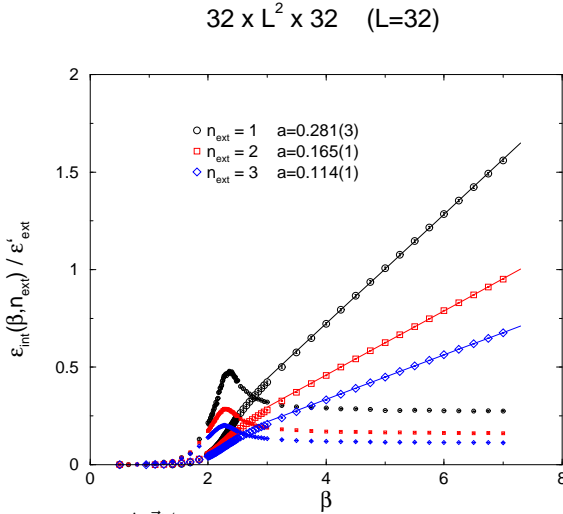}
\end{center}
\vspace{-1.54truecm}
\caption{$\frac{\varepsilon^{\prime}[ \vec{A}^{\text{ext}} ]}
{ \varepsilon^{\prime}_{\text{ext}} }$ versus $\beta$ on a $32^4$ lattice
with the result of the numerical integrations (solid lines).}
\end{figure}

Therefore the  contributions to $\varepsilon^{\prime}[\vec{A}^{\text{ext}}]$  
due to the constrained links
must be subtracted and 
we have to measure the derivative of the ``internal''
energy density $\varepsilon^{\prime}_{\text{int}}$ obtained by
evaluating the contribution to 
$\varepsilon^{\prime}[\vec{A}^{\text{ext}}]$ 
given by the dynamical links.
We numerically integrate  the data and obtain (see Fig.\ref{fig1}) 
for $\beta \gg 1$:
\begin{equation}
\label{aperbeta}
\varepsilon_{\text{int}}(\beta,n_{\text{ext}}) 
\simeq a(n_{\text{ext}}) \,  \frac{H^2}{2}\, 
a(n_{\text{ext}}) \beta  \,,
\end{equation}
where $a(n_{\text{ext}}) \ll 1$ 
to be contrasted\cite{Cea} to the case of U(1) where 
$a(n_{\text{ext}}) \simeq 1$.
%

\section{THE SU(2) VACUUM ENERGY DENSITY}

The data for 
$\varepsilon^{\prime}_{\text{int}}(\beta,n_{\text{ext}}) / 
\varepsilon^{\prime}_{\text{ext}}$
at the perturbative tail and at the peak for various lattice sizes
and values of $n_{\text{ext}}$ can be expressed as a function of the scaling 
variable $x = a_H / L_{\text{eff}}$,
where 
$a_H = \sqrt{2 \pi / gH} = \sqrt{L_1/(2 n_{\text{ext}})}$ is the 
magnetic length  and $ L_{\text{eff}} = \Omega_{\text{int}}^{1/4}$ 
is the lattice effective linear size.

\begin{figure}[t]
\label{fig2}
\begin{center}
\includegraphics[width=7.5cm]{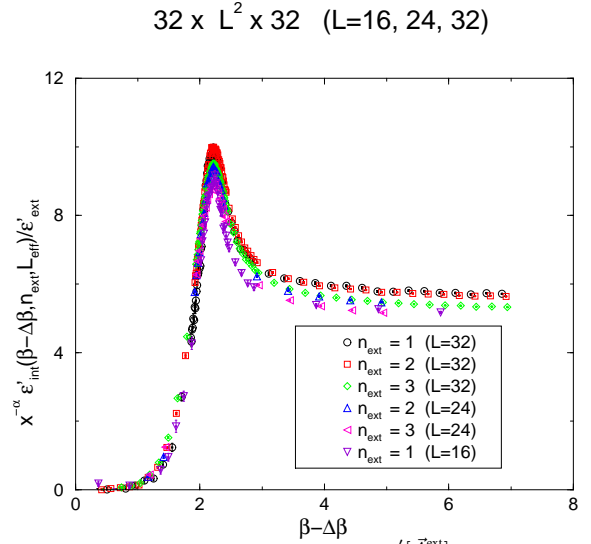}
\end{center}
\vspace{-1.54truecm}
\caption{The universality plot for $\frac{\varepsilon^{\prime}[ \vec{A}^{\text{ext}} ]}
{ \varepsilon^{\prime}_{\text{ext}} }$ versus $\beta-\Delta \beta$.}
\end{figure}

At fixed $\beta$,  we can fit these data to a power-law behavior:
\begin{equation}
\label{powerlaw}
\frac{\varepsilon^{\prime}_{\text{int}}(\beta,n_{\text{ext}})}{\varepsilon^{\prime}_{\text{ext}}} = 
k(\beta) x^\alpha \,.
\end{equation}
This suggests that the lattice data scale according to 
\begin{equation}
\frac{\varepsilon^{\prime}_{\text{int}}(\beta,n_{\text{ext}},L_{\text{eff}})}
{\varepsilon^{\prime}_{\text{ext}}} = 
k(\tilde{\beta}) x^\alpha \,, \quad \alpha=1.5 \,,
\end{equation}
where $\tilde{\beta} = \beta - \Delta \beta$ and $\Delta \beta$ takes care
of the shift of the peak position of $\varepsilon^{\prime}_{\text{int}}$. 
Indeed in Fig.\ref{fig2} we see that all our numerical data 
(for all values of $a_H$ and  $L_{\text{eff}}$) 
can be approximately arranged on the 
scaling curve $k(\tilde{\beta})$ (Eq.(\ref{powerlaw}).
Our numerical data suggests that
\begin{equation}
\label{limitecontinuo}
\lim_{L_{\text{eff}} \to \infty} 
\varepsilon_{\text{int}}(\beta,n_{\text{ext}},L_{\text{eff}}) = 0 \,.
\end{equation} 
in the whole range of $\beta$.
This would imply that in the continuum limit ($L_{\text{eff}} \to \infty$),
$\beta \to \infty$ the SU(2) vacuum screens completely
the external chromomagnetic abelian field, behaving like a 
abelian magnetic condensate medium.
%

\section{FINITE TEMPERATURE}

\begin{figure}[t]
\label{fig3}
\begin{center}
\includegraphics[width=7.5cm]{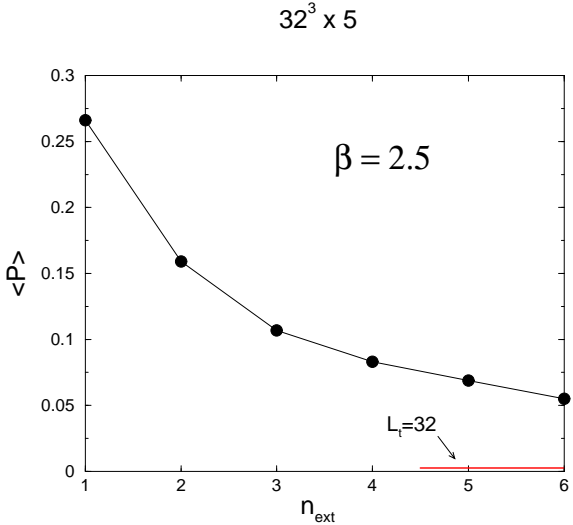}
\end{center}
\vspace{-1.54truecm}
\caption{The temporal Polyakov loop vs. $n_{\text{ext}}$.}
\end{figure}

We also studied the SU(2) gauge system in an external chromomagnetic abelian field
at finite temperature in order to investigate if a connection exists between the 
external chromomagnetic field and confinement.
We looked  at the behavior of the temporal Polyakov loop vs. the
external applied field. 
We considered a SU(2) gauge system at $\beta=2.5$ on a 
$32^3 \times 5$ lattice at 
zero applied external (i.e. $n_{\text{ext}}=0$)
which is in the deconfined phase of finite temperature SU(2).
If the external field strength is increased the expectation value of the Polyakov loop 
is driven towards the value at zero temperature
(see Fig.\ref{fig3}). A similar observation has been reported in Ref.~\cite{Ogilvie1997}. 
On the other hand, if we start with a SU(2) gauge system at zero temperature in a constant 
abelian chromomagnetic background field of fixed strength ($n_{\text{ext}}=1$) and 
increase the temperature the perturbative tail of the 
$\beta$-derivative of the energy density
$\varepsilon^{\prime}_{\text{int}}(\beta,n_{\text{ext}}) / \varepsilon^{\prime}_{\text{ext}}$ 
increases with $1/L_t$  and tends towards the ``classical'' value
$\varepsilon^{\prime}_{\text{int}}(\beta,n_{\text{ext}}) / \varepsilon^{\prime}_{\text{ext}}=1$
We may conclude that increasing the  temperature there is no screening effect 
for the energy density, confirming that the zero-temperature screening of the external field 
is related to the confinement.

Moreover the  information on 
$\varepsilon^{\prime}_{\text{int}}(\beta,n_{\text{ext}}) / \varepsilon^{\prime}_{\text{ext}}$ 
at finite temperature can be used to get an estimate of 
$T_c/\Lambda_{\text{latt}}$.
Our data can be parameterized near the peak as
\begin{equation}
\label{betastar}
\frac{\varepsilon^{\prime}_{\text{int}}(\beta,L_t)}{\varepsilon^{\prime}_{\text{ext}}} =
\frac{a_1(L_t)}{a_2(L_t) \, \left[ \beta - \beta^{*}(L_t) \right]^2 + 1} \,.
\end{equation}
So that we get an estimate of the deconfinement temperature (see Fig.4):
\begin{eqnarray}
\label{temperature}
\lefteqn{
\frac{T_c}{\Lambda_{\text{latt}}} = \frac{1}{L_t} \, \frac{1}{f(\beta^{*}(L_t))}
= 27.12 (4.04)}  \nonumber \\
\lefteqn{
f(\beta)=\left( \frac{11}{6 \pi^2} \, \frac{1}{\beta} \right)^{-\frac{51}{121}} \, 
\exp \left( - \frac{3 \pi^2}{11} \beta \right) } \,.
\end{eqnarray}
Our result is consistent with previous studies~\cite{Fingberg1993}.
\begin{figure}[t]
\label{fig4}
\begin{center}
\includegraphics[width=7.5cm]{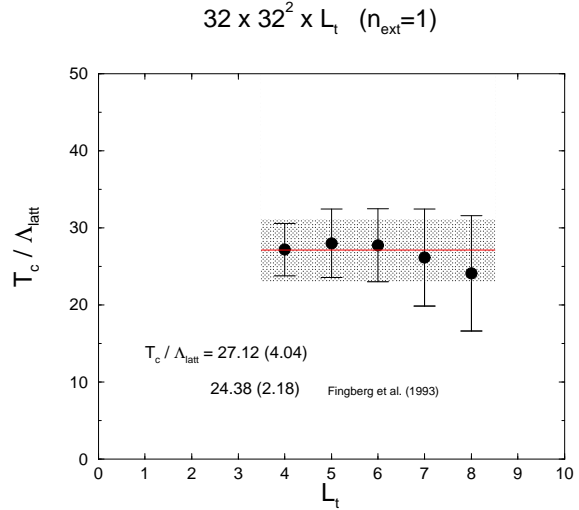}
\end{center}
\vspace{-1.54truecm}
\caption{$T_c/\Lambda_{\text{latt}}$ vs. $L_t$.}
\end{figure}
%
%

\section{CONCLUSIONS}

We have studied the non-perturbative dynamics of the SU(2) l.g.t. vacuum
in an external abelian chromomagnetic field, 
by means of a gauge invariant 
effective action  defined using the lattice Schr\"odinger functional.

Our numerical results suggests that in the continuum limit 
($L_{\text{eff}} \to \infty$, $\beta \to \infty$) the energy density
$\varepsilon_{\text{int}}(\beta,L_{\text{eff}},n_{\text{ext}}) \to 0$, 
i.e.:
the SU(2) vacuum screens completely the external abelian chromomagnetic field and
behaves like an abelian magnetic condensate in agreement with the 
dual superconductor scenario for the color confinement.

At finite temperature it seems that confinement is restored by increasing the
strength of the external applied field. On the other hand the zero temperature screening of the
external field is removed by increasing the temperature.




\begin{thebibliography}{99}
\bibitem{Rossi1980} G. C. Rossi and M. Testa, Nucl. Phys. {\bf B163} (1980) 109; {\it ibid.}
{\bf B176} (1980) 477.
\bibitem{Luscher1992} M. L\"uscher, R. Narayanan, P. Weisz, and U.
Wolff, Nucl. Phys. {\bf B384} (1992) 168; 
M. L\"uscher and   P. Weisz,   Nucl. Phys. {\bf B452} (1995) 213.
\bibitem{Cea} P. Cea, L. Cosmai, and A. D. Polosa, Phys. Lett. {\bf B392} (1997) 177;
Phys. Lett. {\bf B397} (1997) 229; P. Cea, L. Cosmai,  Mod. Phys. Lett. {\bf A13} (1998) 861.
\bibitem{Ogilvie1997}  M. Ogilvie,  Nucl. Phys. B (Proc.Suppl) {\bf} 63 (1998) 430.
\bibitem{Fingberg1993} J. Fingberg, U. Heller, F. Karsch, Nucl. Phys. {\bf B392} (1993) 493. 
\end{thebibliography}
\end{document}